\documentclass[aps,prl,twocolumn,showpacs,superscriptaddress,groupedaddress,nofootinbib]{revtex4}  
\usepackage{graphicx}  
\usepackage{dcolumn}   
\usepackage{bm}        
\usepackage{amssymb}   
\usepackage{amsmath,amsthm,latexsym,amssymb,amsfonts,epsfig}
\usepackage{float}
\usepackage{adjustbox}
\usepackage{multirow}
\usepackage{tabularx}
\usepackage{mathtools}
\usepackage{float}
\usepackage[latin9]{inputenc}
\usepackage[usenames,dvipsnames]{xcolor}
\newcommand{\be}{\begin{equation}}
\newcommand{\ee}{\end{equation}}
\newcommand{\bea}{\begin{eqnarray}}
\newcommand{\eea}{\end{eqnarray}}

\begin{document}

\title{More on Emergent Dark Energy from Unparticles}

\author{Micha{\l} Artymowski}

\affiliation{Cardinal Stefan Wyszy{\'n}ski University, College of Science, Department of Mathematics and Natural Sciences, Dewajtis 5, 01-815 Warsaw, Poland}
\author{Ido Ben-Dayan}
\affiliation{Physics Department, Ariel University, Ariel 4070000, Israel}
\affiliation{Berkeley Center for Cosmological Physics, University of California, Berkeley, CA 94720, USA}
\affiliation{Department of Physics, University of California, Berkeley, CA 94720, USA}
\author{Utkarsh Kumar}
\affiliation{Physics Department, Ariel University, Ariel 4070000, Israel}
\date{\today}

\begin{abstract}
In a recent paper \cite{Artymowski:2020zwy} we suggested the possibility that the present acceleration of the Universe is due to thermodynamical behavior of unparticles. The model is free of scalar fields, modified gravity,a Cosmological Constant (CC), the coincidence problem, initial conditions problem and possesses interesting distinct predictions regarding the equation of state of Dark Energy, the growth rate and the number of relativistic degrees of freedom at BBN and CMB decoupling. In this work we relate to a recent paper \cite{Abchouyeh:2021wey}, which  discusses a similar setup of unparticles with and without a CC as an external source of late-time acceleration. The authors have shown how such a model is inconsistent with the data. We show that these claims are viable only in a particular part of the parameter space and that model \cite{Artymowski:2020zwy} stands tall. We further suggest a consistency condition in terms of observables. We then fit publicly available supernovae data to derive the expected Hubble parameter and constrain the parameters of the model. 

\end{abstract}

\maketitle
\section{Introduction}
An extensive analysis studying the different phases of unparticles \cite{Georgi:2007ek,Grzadkowski:2008xi,Tuominen:2012qu} in cosmology has been carried out in \cite{Artymowski:2019cdg}. The analysis discovered new cyclic and bouncing cosmological models, as well as standard hot Big Bang scenarios, and a possible Dark Energy (DE) model. The Dark Energy model was extensively studied in \cite{Artymowski:2020zwy} \footnote{{Different approach to DE models motivated by unparticles were also studied in \cite{Dai:2009mq,Diaz-Barron:2019uzd}}. For more details on DE see \cite{Copeland:2006wr,Bamba:2012cp}}. The outcome of the study was that unparticles act as radiation in the early universe while at late times act as effective DE and asymptote to a Cosmological Constant (CC) for a certain range of parameters. The model has distinct predictions of enhanced number of relativistic degrees of freedom, $N_{eff}$, special redshift dependence of the equation of state parameter $w_u(z)$, and some small deviations from the growth history of the $\Lambda$CDM model. It was further suggested that the unparticles emergent DE model could reduce the Hubble tension. The main theoretical novelty is that the DE behavior is due to an emergent collective behavior that is temperature dependent and is not based on scalar fields or modified gravity.
In recent work, the question of unparticles as a relevant fluid in the late-time universe and the stability of such a model was discussed \cite{Abchouyeh:2021wey}, claiming that unparticles cannot govern the late-time universe and therefore cannot act as a DE model or reduce the Hubble tension \footnote{{The so called Hubble tension is the growing discrepancy between various inferences or measurements of the Hubble parameter. Most notably, there seems to be a discrepancy between early universe and late universe probes. In this manuscript, we quantitatively mean the discrepancy planck 2018 value of $H_0=67.4$km/Mpc/sec \cite{Aghanim:2018eyx}. and the cepheid based SN measurement of $74.3$ km/Mpc/sec \cite{Riess:2019cxk} and the quoted standard deviations.}. This conclusion is valid only in a certain range of parameters. In \cite{Artymowski:2020zwy} and in this manuscript, we use a different range of parameters.} We demonstrate explicitly the stability of the DE unparticle model. We derive a new consistency condition of the model between $N_{eff}$ and $w_u(z)$. Finally, we { perform a likelihood analysis to $\Lambda$CDM and to our unparticles DE model using }publicly available Supernovae data.

\section{Allowed range of parameters}

Consider the flat Friedmann-Lemaitre-Robertson-Walker (FLRW) metric in a universe filled with unparticles, radiation and matter. The system is described by the following line element and equations of motion (not all of them independent):
 \begin{eqnarray} 
 ds^{2} &=& -dt^2 + a^{2}(t) \, d \Vec{x}^2 \label{eq:dS}\\
 3 H^2 &=& \rho_r + \rho_m + \rho_u , \label{eq:1Friedmann}\\
 \dot{H} &=& -\frac{1}{2}\left( \frac{4}{3} \, \rho_r + \rho_m + \rho_u + p_u \right), \label{eq:2Friedmann}\\
 \dot{\rho}_i &+& 3 H \left(\rho_i+p_i\right) = 0. \label{eq:continuity}
 \end{eqnarray}
 where $"i"$ stands for radiation, matter and unparticles. 
 
Our starting point is that for temperatures {of unparticles} lower than some cut-off scale the energy density and pressure of unparticles are described by the following equations:
\begin{eqnarray} \label{eq:basic}
\rho_u &=& \sigma T^4 + A\left(1 + \frac{3}{\delta}\right) T^{4 + \delta}\equiv\sigma T_c^4 y^4 \left(1-\frac{4 (\delta +3) y^{\delta }}{3 (\delta
   +4)}\right)\cr
p_u &=& \frac{1}{3}\sigma T^4 + \frac{A}{\delta}T^{4 + \delta}\equiv \frac{\sigma}{3}  T_c^4  y^4 \left(1-\frac{4 y^{\delta }}{\delta
   +4}\right),
\end{eqnarray}
where {$\delta = a + \gamma$, $a>0$ is a constant that determines the $\beta$ function around the IR fixed point, $\gamma$} is an anomalous dimension, {$A$ \footnote{In \cite{Artymowski:2020zwy}, we used the parameter $B\equiv A(1+3/\delta)$, to simplify the analysis. We will use $B$ and $y$ from now on.} is defined as $\frac{a u b}{2g*}$ where $u$ is an integration constant, and $g_*$ the value of the coupling constant at the fixed point of the Bank-Zaks theory. More details can be found in \cite{Artymowski:2019cdg} and references therein}. $\sigma>0$ is related to the number of degrees of freedom, and that the temperature of unparticles $T$ can in principle take any positive semi-definite value. {We want to emphasize that $T$ is the temperature of unparticles, and \emph{not} the temperature of radiation, which satisfies a separate continuity equation. In addition, one assumes that unparticles and SM particles are coupled only in the high energy regime, while in the considered range of temperatures they are fully decoupled (except for gravity of course). Thus, unparticles may be consider as a dark fluid.} For convenience, in the second equality we have switched to dimensionless temperature $y\equiv T/T_c$, where $T_c = \left(\frac{4(\delta+3)}{3(\delta+4)}(\frac{-\sigma}{B})\right)^{1/\delta}$ is a critical temperature where $\rho_u+p_u=0$. One can use the dimensionless temperature $y$ only if $T_c$ and $\rho_u(T_c)$ are real and positive, which restricts the allowed parameter space to $B<0$ and $-3<\delta<0$, which, using $\delta=2(d_{\mathcal{U}}-1)$ \cite{Abchouyeh:2021wey}, corresponds to the {scaling} dimension $-1/2<d_{\mathcal{U}}<1$. The equation of state (EOS) parameter of unparticles is then defined as 
\be
w_u=\frac{1}{3}\frac{\sigma+\frac{3}{\delta+3} B \,T^{\delta}}{\sigma +B \,T^{\delta}}
 \ee
Hence, a simple estimate shows that $w_u$ interpolates between $w_u=1/3$ and $w_u=(\delta+3)^{-1}$ depending on the temperature of unparticles. This estimate does not take into account how the temperature of unparticles is evolving. However, if it is correct then one may simply approximate $w_u$ as one of these aforementioned values in the late universe and consider its implications. In such a case, it is also clear that for a successful DE model one should have $\delta\sim -4$.
 
{\cite{Abchouyeh:2021wey} discusses unparticle cosmology for $B > 0$, concluding this parameter regime to offer no improvement over $\Lambda$CDM in resolving the $H_0$ tension, shown in their Fig. 10.} However, the Authors of \cite{Abchouyeh:2021wey} have investigated only a fraction of possible range of temperatures of unparticles. As we have shown in \cite{Artymowski:2020zwy} $B <0$ results in a different behavior and a valid DE model with implications for several observables in Cosmology. Specifically, the approximation $w_u\simeq(\delta+3)^{-1}$ is insufficient for understanding the dynamics of the unparticles DE model.

For a valid model we require an expanding universe $H>0$, that the energy density of unparticles $\rho_u\geq0$ is always positive semi-definite, the Null Energy Condition (NEC) $\rho_u+p_u\geq 0$ and that the temperature is decreasing with time $dT/dt<0$ which from the continuity equation and NEC is equivalent to $d\rho_u/dT\geq 0$ and is equivalent to $da/dT <0$, since $\dot{a} = \frac{da}{dT}\frac{dT}{dt} > 0$. Hence,
\begin{eqnarray} \label{eq:range}
\rho_u &=& \sigma T^4\left(1 + \frac{B}{\sigma}\, T^{\delta}\right) \geq 0,  \\
\rho_u+p_u &=& \sigma T^4\left(\frac{4}{3} + \frac{B}{\sigma}\frac{\delta+4}{\delta+3}\,T^{ \delta}\right)\geq 0, \label{eq:NEC}\\
\frac{d\rho_u}{dT}&=&\sigma T^3\left(4 + \frac{B}{\sigma}(\delta+4)\,T^{ \delta}\right)\geq 0 \label{eq:rhoT}
\end{eqnarray}
For $B>0,\,\delta>-3$ the above equations are fulfilled for all $T\geq 0$, i.e. they are describing a fluid in an expanding universe with positive energy density, that does not violate the NEC and that its temperature is positive and decreasing with time and with the expansion of the universe as one would expect. Any other combination of $B,\delta$ would result in a violation of one of the inequalities at some finite temperature, which in principle does not need to be a problem as long as the physical evolution of unparticles never leads to values of $T$ that would violate (\ref{eq:range}-\ref{eq:rhoT}). Since the Authors of \cite{Abchouyeh:2021wey} are considering the $T \to 0$ regime, one must remember that the allowed parameter space for their analysis shall also be restricted to $B>0$ and $\delta> -3$, since for negative $\delta$, $B>0$ and sufficiently small temperatures \eqref{eq:NEC} requires $\delta<-4$ or $\delta>-3$ and the \eqref{eq:rhoT} requires $\delta>-4$. Note that assuming $B<0$ would automatically lead to $\rho_u<0$ for $\delta<0$ and $T\to0$. To conclude, the parameter space considered by the Authors of \cite{Abchouyeh:2021wey} (namely $\delta\in(-6,1)$) is too broad for a valid model of DE.\footnote{{The $\delta=-4$ case is not really a DE model, but simply the trivial case of radiation and a CC at any $T$.}}. Nevertheless, as we will show below, the shortcomings of their analysis are not coming from a wrong set of $\delta$.

\begin{figure}[]
\includegraphics[width=5.2cm, height=4.8cm]{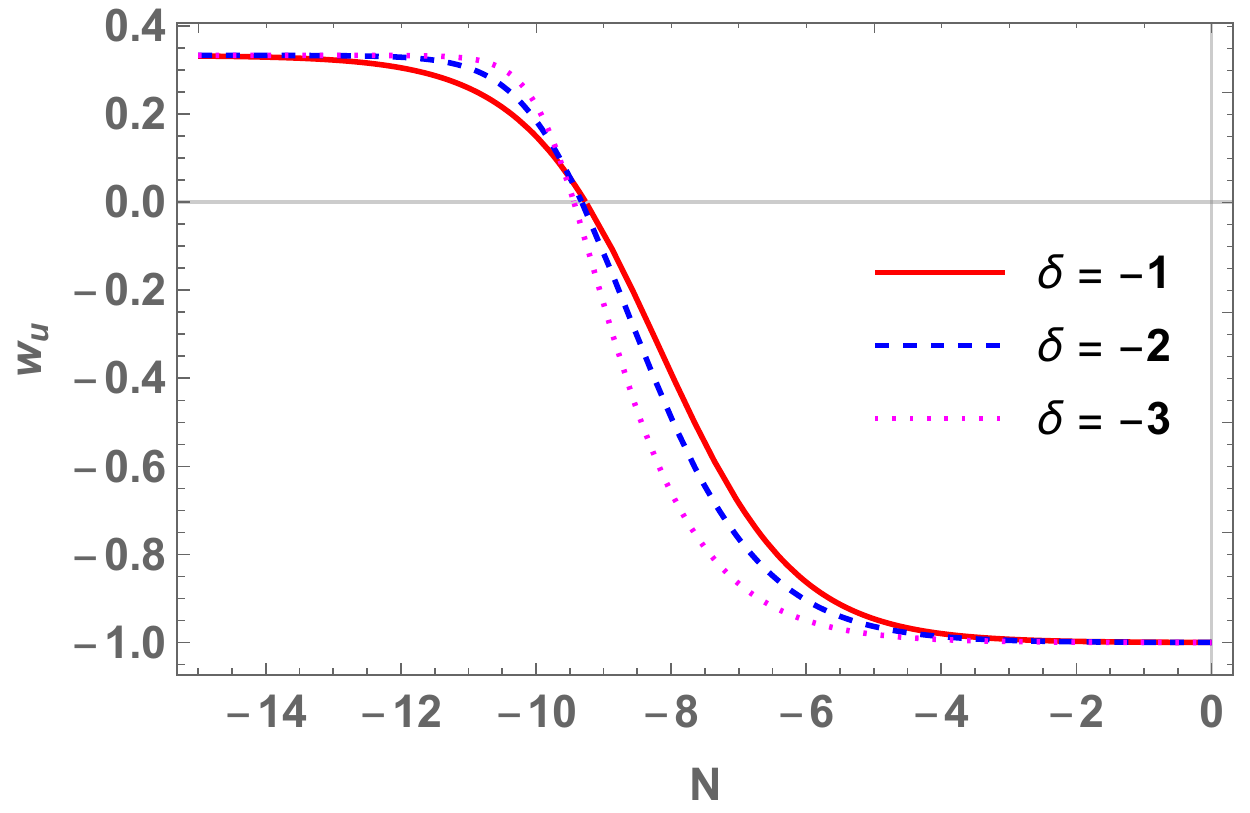}
\caption{\it The equation of state of unparticles $w_u$ as a function of e-folds, $N=\ln a(t)$, for $\delta=-1,-2,-3$. The unparticles have reached a CC behavior $w_u\simeq -1$.} 
\label{fig:tracker}
\end{figure}

{One can take the limit where} $w_u\sim (\delta+3)^{-1}$ and ask how such an approximation gives a DE behavior. We do not neglect the fact that for some part of the parameter space one can reach $T\to0$ limit, which makes the analysis presented in \cite{Abchouyeh:2021wey} valid. Nevertheless, there in an important region of the parameter space {$B<0$}, discussed in \cite{Artymowski:2019cdg,Artymowski:2020zwy}, which does not allow $T$ to reach arbitrarily low temperatures {where $w_u\sim (\delta+3)^{-1}$ is erroneous}. The important point is that solving the system of equations one sometimes cannot reach $T=0$ through the physical evolution of the Universe. This conclusion comes from the fact that for $B<0$ and $-3<\delta<0$ the continuity equation of unparticles gives
\begin{equation}
a(y) \propto y^{-1}\left|y^\delta -1\right|^{-1/3} \, , 
\end{equation}
Note that $T=T_c$ is a pole of the scale factor and thus one cannot cross it through standard physical evolution. This conclusion does not depend on the possible existence of additional fluids, like dust or radiation, since the continuity equation for unparticles remains valid in the multi-component Universe. Starting from a high enough temperature there is always a finite temperature $T=T_c$ where $\dot{H} \supset \rho_u+p_u\rightarrow 0$, and the unparticles temperature asymptotes to $T_c$ from above. Approaching $T_c$ unparticles effectively act as a CC with $w_u\simeq -1$. An example of the evolution of $w_u$ is presented in Fig. \ref{fig:tracker}.

We wish to stress that this result is not due to some additional limitation on dynamics. It is simply the solution of the equations of motion. In $T \simeq T_c$ regime one finds $\rho_{DE,0}\equiv \rho_u \simeq \rho_u(T_c) = \frac{-\delta\sigma}{3(\delta+4)}T_c^4$, which gives $B$ in terms of the present day DE density. Even though we consider $B<0$, one never violates (\ref{eq:range}-\ref{eq:rhoT}), since $T$ can never reach the $T<T_c$ regime. The whole evolution of the Universe happens outside of $T\to0$ region, for which $w_u\sim (\delta+3)^{-1}$ would be valid. We also want to emphasize that the low-energy DE behaviour of unparticles appears for all $\delta\in(-3,0)$. This stands in a stark contrast to the result of \cite{Abchouyeh:2021wey}, where the CC-like behaviour of unparticles happens only for $\delta\simeq -4$. {This CC-like behaviour can be obtained by using $w_u \sim (\delta+3)^{-1}$ valid to $B>0$.}

To understand whether unparticles temperature asymptote to $T_c$ or not, let us note that in the expanding Universe one finds $H>0$, so from $H=\frac{da}{dt}\frac{1}{a} \propto \frac{da}{dt} = \frac{da}{dT}\frac{dT}{dt} = \frac{da}{dy}\frac{dy}{dt}$. Thus, one finds $\frac{dy}{dt} >0$ or $\frac{dy}{dt} < 0$ for $\frac{da}{dy} >0 $ or $\frac{da}{dy} <0 $ respectively. The $\frac{dy}{da} >0$ condition corresponds to
\begin{equation}
\frac{da}{dy} > 0 \Leftrightarrow T \in \left(\left(\frac{3}{3+\delta}\right)^{1/\delta}T_c,T_c \right)\, . \label{eq:yinequality}
\end{equation}
Eq. \eqref{eq:yinequality} is valid regardless of the possible presence of dust or radiation. This means that $T$ will asymptotically approach $T_c$ for $T > \left(\frac{3}{3+\delta}\right)^{1/\delta}T_c$ no matter what currently dominates the Universe. $T< \left(\frac{3}{3+\delta}\right)^{1/\delta}T_c$ is the only range, for which $T<T_c$ and the temperature decreases with the scale factor. Thus, it is the only range in which one can reach $T\to 0$ limit for $B<0$ and $-3<\delta<0$.

\begin{figure}[h]
\centering
 \begin{adjustbox}{center}
\includegraphics[width=4.8cm, height=4.5cm]{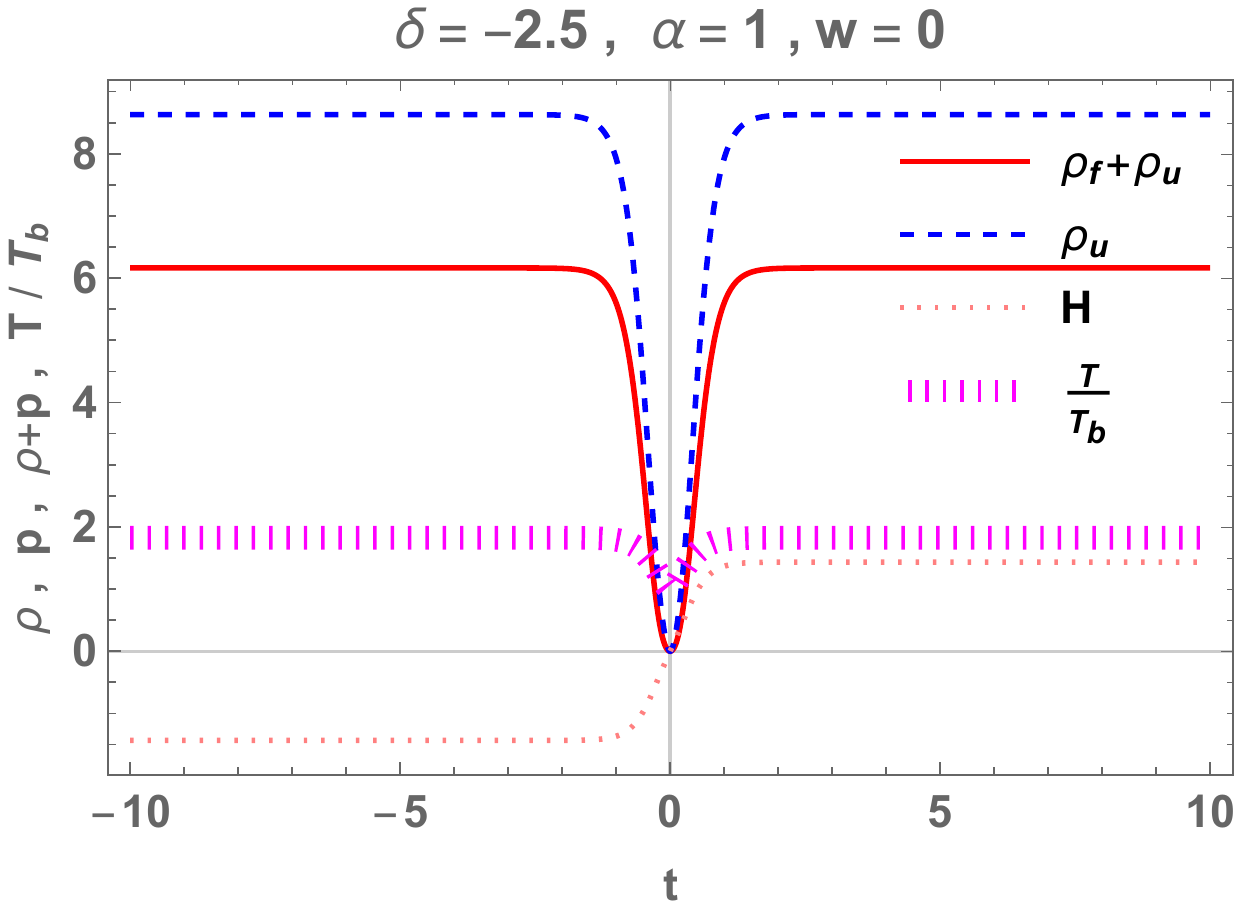} 
\includegraphics[width=4.8cm, height=4.5cm]{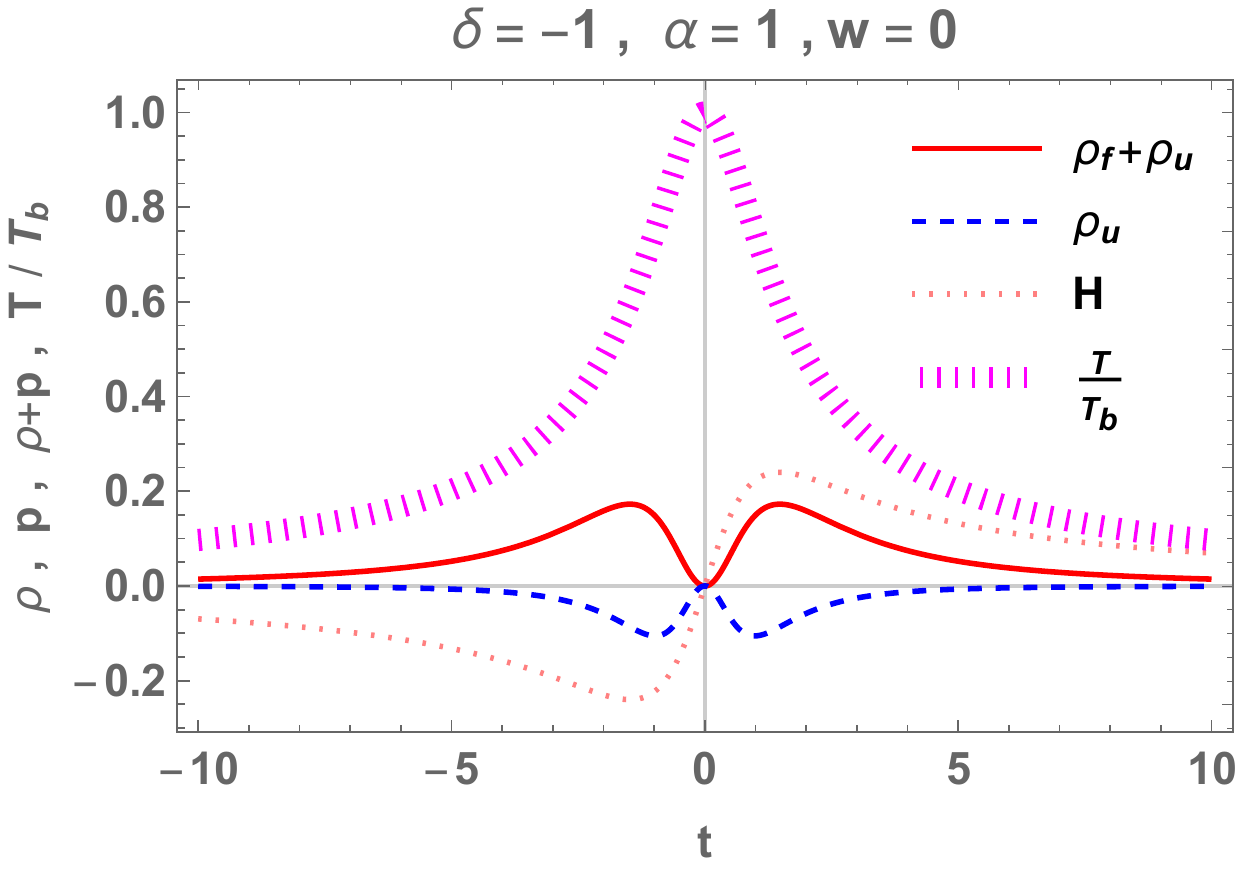}
\end{adjustbox}
\caption{\it Left and right panels represent two possible scenarios for the evolution of the Universe in the $T<T_c$ regime. We assume the Universe filled with unparticles and dust and $\alpha = \rho_{f0}/(\sigma T_b^4)$. Reproduced from \cite{Artymowski:2019cdg}.} 
\label{fig:bounces}
\end{figure}

The case of $T<T_c$ has been discussed in \cite{Artymowski:2019cdg}. For the discussed range of parameters, $B<0$ and $-3<\delta<0$, and for the Universe filled with unparticles and some perfect fluid, one finds two possible scenarios. The first option is the de Sitter bounce, which means that the Universe collapses exponentially, which is followed by a non-singular bounce and an exponential expansion of the scale factor. An example is presented in the left panel of Fig. \ref{fig:bounces}, where $T_b = (-\sigma/B)^{1/\delta}$ is the temperature at the bounce. The range between $T_b$ and $T_c$ lies within \eqref{eq:yinequality}, thus one finds $dT/da>0$. Hence, the unparticle temperature will \textit{increase} with the expansion of the universe, again asymptoting to $T=T_c$. $T$ obtains its minimum at the bounce and it can reach $T_{min}\simeq 0$ for $\delta\gtrsim -3$. In such a case the $w_u\sim(\delta+3)^{-1}$ is valid only in the close vicinity of the bounce, but not in the late-time evolution of the Universe. This could have interesting implications for some previous epochs of the universe, but it cannot act as a DE model.

The other option for the evolution of the Universe with $T<T_c$ is the single bounce also discussed in \cite{Artymowski:2019cdg}. An example is presented in the right panel of Fig. \ref{fig:bounces}. In such a case one starts with a contracting Universe, followed by the bounce and decelerated expansion. The bounce appears at $T = T_b = (-\sigma/B)^{1/\delta}$, and since the Eq. \eqref{eq:yinequality} is not satisfied, $T_b$ is the maximal temperature of the Universe. In this scenario the allowed range of temperature is $T\in (0,T_b)$ and therefore one can reach $w_u \simeq (\delta + 3)^{-1}$. Nevertheless, this range of temperatures is still unfit for any DE model building, because $\rho_u < 0$ throughout the whole evolution. 

To conclude, the only possible option for the unparticle DE is $-3<\delta<0$, $B<0$, and initial $T>T_c$. In such a case one obtains {late-time} CC-like behavior of unparticles for all $\delta\in(-3,0)$. For the same range of parameters and the initial temperature $T<T_c$ one always obtains a bouncing scenario, which may be a dS bounce with $T$ approaching $T_c$ or a single bounce with $\rho_u<0$. Only in the latter case one can reach $T\to0$. The analysis done in \cite{Abchouyeh:2021wey} is valid only for $B>0, -3<\delta<0$, which excludes the existence of $T_c$ and late-times behavior of unparticles.

The $T\to0$ limit the Authors have chosen is designed to fail in the context of DE model building, since for {a valid model} one needs to reach an almost constant energy density, which will never happen if $T \to 0$.

\section{Stability Analysis of Unparticles Dark Energy} 

\begin{figure}[h]
\centering
 \begin{adjustbox}{center}
\includegraphics[width=4.5cm, height=3.8cm]{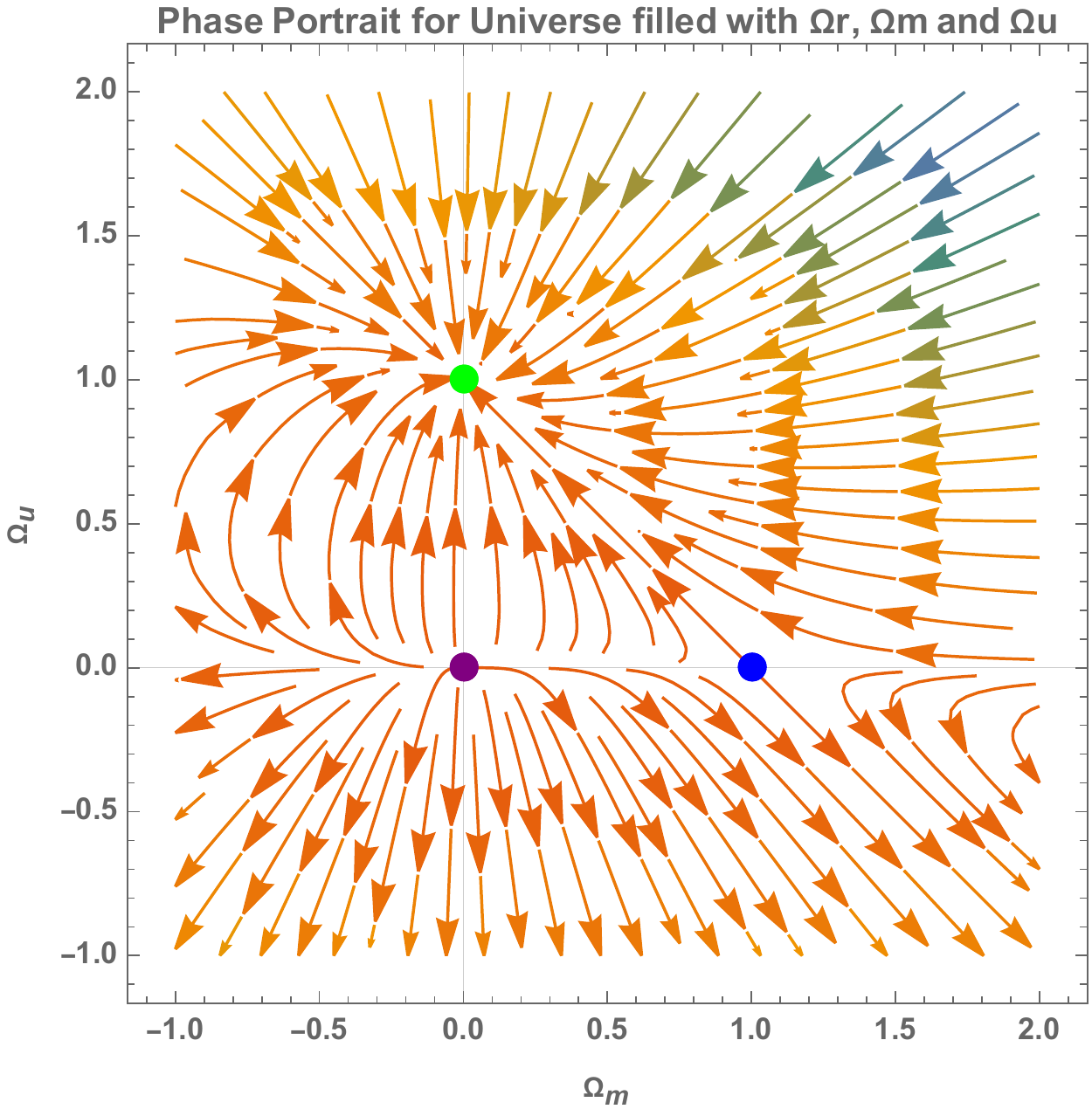} 
\includegraphics[width=4.5cm, height=3.8cm]{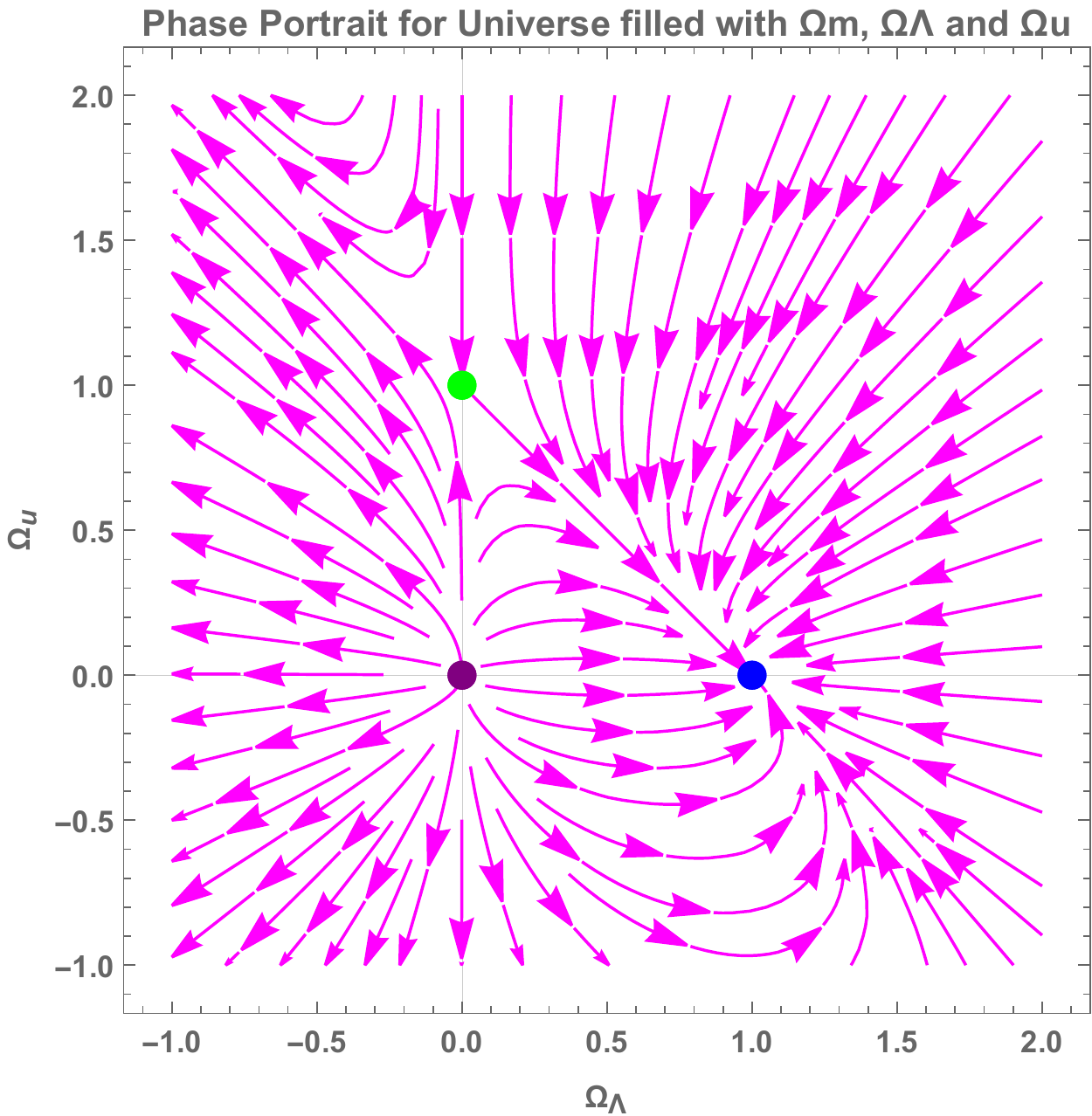}
\end{adjustbox}
\caption{\it Left panel: Phase portrait for a Universe filled with radiation, matter and unparticles. Green, blue and purple points represent the unparticles, matter and radiation fixed points respectively. Unparticles with $w_u=-1$ is the stable fixed point. \it Right panel : Phase portrait for a Universe filled with  matter, cosmological constant and unparticles. Green, blue and purple points represent the unparticles, cosmological constant and matter fixed points respectively. {We chose $w_u=-0.5$. The CC with $w = -1$ will ultimately dominate until unparticles strictly become a CC in the infinite future, where the phase portrait will have a degenerate straight stability line connecting the points of unparticle domination and CC domination. This straight line will} correspond to $\Omega_{\Lambda}+\Omega_u=1$.} 
\label{fig:phase}
\end{figure}

After establishing the allowed parameter range and the temporal behavior of unparticles, i.e. that starting from a large enough temperature they asymptote to $T=T_c$ and act as a DE, let us discuss the stability of the solution. We do not claim that the stability analysis done in \cite{Abchouyeh:2021wey} had some mathematical error, but that it considered a different range of parameters and initial conditions. We therefore want to show the stability of the Universe for the realistic unparticle model of DE, where $w_u$ asymptotes to $w_u\downarrow -1$. Consider the background evolution of a  Universe filled with multiple fluids, such as radiation, dust, cosmological constant and unparticles. Strictly speaking the analysis presented here is valid for constant or slowly varying EOS, $w$. For a full stability analysis, one must take into account the time variation of $w$, spatial dependence of perturbations and show that $0<c_s^2<1$ to avoid superluminality. We have taken this into account in \cite{Artymowski:2020zwy}, when calculating the effects of the unparticle DE model on the growth index $\gamma$ and $f\sigma_8$. We got a stable system with unparticles acting as DE. Nevertheless, it is worthwhile to discuss the stability analysis of the background only as well \cite{An:2015mvw,Paliathanasis:2015gga}.
First, let us consider a single fluid. It is instructive because it is almost always a good approximation to cosmology since generically one fluid dominates over all the others. Using the two Friedmann equations we get
\be
\dot{H}=-\frac{3}{2}(1+w)H^2
\ee
Obviously, any perturbation $\delta H(t)$ will die out with time as long as $w>-1$, be it unparticles, radiation, matter or any other energy content. 
Let us now analyze the multi-component case. In order to perform the stability analysis, we choose to work with dimensionless variables defined as:
\begin{eqnarray*}
\Omega_i = \frac{\rho_i}{3\, H^{2}}, \quad
 N = \ln a(t).
\end{eqnarray*}
Substituting $\Omega_i$ and $ N$ in the continuity equations \eqref{eq:continuity} and after simplifying one arrives at the following system
\begin{eqnarray}
\Omega_i' &=& - 3 \left( 1 + w_i + \frac{2}{3}\, \frac{H'}{H}\right)\Omega_i 
\end{eqnarray}
Here $' = \frac{d}{d N}$, and $\frac{H'}{H} = -\frac{3}{2} \left( \frac{4}{3}\, \Omega_r + \Omega_m + \left(1 + w_u \right)\, \Omega_u\right)$ for a universe filled with unparticles, radiation and matter. Substituting back $\frac{H'}{H}$ into the continuity equations  and removing $\Omega_r$ by using $\Omega_r = 1 - \Omega_m - \Omega_u$ results in two equations $i=m,u$: 
\begin{eqnarray}
\Omega_i' &=&   \left(1 -3 w_i - \Omega_m - \left(1 -3 w_u\right) \Omega_u\right)\, \Omega_i . \label{eq:contgen}
\end{eqnarray}
Fixed points of the dynamical system are calculated by demanding $\Omega_i' = 0$ yielding:
\begin{eqnarray*}
\left(\Omega_m,\Omega_u \right)
&\rightarrow& \left(1,0\right) \rightarrow \text{Matter Domination, (MD)}, \\
&\rightarrow& \left(0,1\right) \rightarrow\text {Unparticle Domination, (UD)}.
\end{eqnarray*}
$\Omega_r = 1$ corresponds to $(0,0)$ and it is unstable, as shown in the left panel of Fig. \ref{fig:phase}.  Stability of each fixed point is determined by calculating the eigenvalues of Jacobian matrix $M_{ij}$, given as 
\begin{eqnarray}
M_{ij} &=& \begin{bmatrix}
\frac{d \Omega_m'}{d \Omega_m} & \frac{d \Omega_m'}{d \Omega_u} \\
\frac{d \Omega_u'}{d \Omega_m} & \frac{d \Omega_u'}{d \Omega_u} 
\end{bmatrix}
\end{eqnarray}
 For the MD fixed point the eigenvalues are $-1 $ and $-3\,w_u$ while for the UD fixed point they are $3\,w_u$ and $3\,w_u -1$. If $w_u\sim (\delta+3)^{-1}$ with $-3<\delta<0$ then $1/3>w_u>0$ so MD is the stable point and UD is unstable as derived in \cite{Abchouyeh:2021wey}. But as stressed, this is {not the case if $B<0$. If $B<0$ then starting} at $T>T_c$, ($w_u$) evolves from $1/3$ to $-1$, hence, once $w_u<0$ UD becomes the stable point and MD the unstable. $w_u$ continues to evolve and asymptotes to $w_u\downarrow -1$ yielding a stable UD  dS era that continues forever. {Let us stress that this conclusion is based on using \eqref{eq:basic} at all times. It may be changed if higher order corrections to the beta function are calculated or if one means a different microscopic model as in \cite{Dai:2009mq,Diaz-Barron:2019uzd}.} 
 
Let us also consider a true CC along with matter and unparticles. In this case we are dropping  the radiation part as it is negligible during late-times, though as we shall see we can apply the analysis to any number of fluids. Following the procedure discussed above, fixed points are given as 
 \begin{eqnarray*}
\left(\Omega_m, \Omega_u,\Omega_{\Lambda} \right)
&\rightarrow& \left(1,0,0\right) \rightarrow \text{MD}, \\
&\rightarrow& \left(0,1,0\right) \rightarrow \text{UD}, \\
&\rightarrow& \left(0,0,1\right) \rightarrow\text {CCD},
\end{eqnarray*}
The eigenvalues for matter, cosmological constant and unparticles fixed points are  $\left(3, -3\,w_u\,,3 \right)$, $\left(0\,,\, -3\left( 1 + w_u\right)\,, -3 \right)$ and $\left(3\left(1 + w_u\right)\,,\, 3 \, w_u ,3\left(1 + w_u\right)\right)$ respectively. As before the stability depends on $w_u$. As unparticles asymptote to $w_u\downarrow -1$, they will be equivalent to the CC in terms of stability. Figure (\ref{fig:phase}) shows the phase portrait of both cases discussed.

The above analysis can be generalized to any number of fluids with general, even time-dependent equation of states in a straightforward manner. 
Starting from the continuity equations, the second Friedmann equation and changing the time variable to e-folds, one reaches the following system for any number of fluids:
\be
\Omega_i'=-3\sum_{j\neq i}(w_i-w_j)\Omega_i\Omega_j=-3\sum_{j}(w_i-w_j)\Omega_i\Omega_j
\ee
where $w_j$ is the EOS of the jth fluid and can have arbitrary time dependence.
We can then derive the Jacobian
\be
\frac{\partial \Omega_i'}{\partial \Omega_k}=-3\left[ \sum_j(w_i-w_j)\Omega_j\delta_{ik}+(w_i-w_k)\Omega_i\right]\label{eq:stabilityM}
\ee
where we kept the sum explicit as not to confuse the summed and unsummed indices.
We are interested in fixed points where eventually one fluid dominates the energy density if the universe. Without loss of generality, let us choose this fluid to be the first one, i.e. $\Omega_1=1,\, \Omega_{i\neq1}=0$. Substituting the fixed point into the Jacobian, the $i=k=1$ entry will be zero, and except for the first line, the rest of the Jacobian is diagonal with $-3(w_i-w_1)$. Hence the eigenvalues of the Jacobian are $-3(w_i-w_1)$ for all $i$. The zero eigenvalue is because we can remove one fluid completely using $1=\sum_i \Omega_i$. For stability, the other eigenvalues require $w_i>w_1 \quad\forall i$, i.e. as long as $w_1$ is the lowest equation of state parameter it will govern the late time behavior of the universe and will be a stable fixed point. Finally, one would still like $w_1\geq -1$. Otherwise, if this fluid dominates, as it should at late times, it will be unstable as can be understood from the single fluid case. These results are also obvious from directly solving the equations of motions for different fluids. Any other result would have been disconcerting as it would mean that various fluids with different equations of state could easily destabilize the universe and falsify the whole idea of FLRW Cosmology. Hence, if unparticles have the lowest EOS with $w_u\geq -1$ they will be a stable point, while if they are not, then they may not. The analysis above is completely general, and the conclusions about stability may only change if some large spatial gradients, ghosts or superluminality appear.  

In \cite{Abchouyeh:2021wey}, {since they limited themselves to $B>0$}, they considered two scenarios with 3 fluids. First, matter, unparticles and a CC with EOSs $0,(\delta+3)^{-1},-1$ respectively. Second, matter, unparticles and radiation, with EOSs $0,(\delta+3)^{-1},1/3$. In both scenarios with $-3<\delta<0$ there is a fluid or more with an equation of state lower than the unparticles (matter and CC), which is why they got that unparticle domination is an unstable fixed point. {Here, as demonstrated, since $B<0$ it will be incorrect to simply approximate $w_u=(\delta+3)^{-1}$}. Rather, unparticles have a time dependent equation of state that is affected by the dynamics of the universe. If we start from $T>T_c$ then the temperature of unparticles asymptote to $T_c$ yielding $w_u\downarrow -1$.
Hence, for a universe filled with radiation, matter and unparticles at some point in time unparticles will become a stable fixed point and will dominate the energy density of the universe. If there is a bare CC, then the unparticles will asymptote to a CC and become indistinguishable from it in the asymptotic infinite future, so this is still a stable solution.

Finally, perturbations from emergent DE model of unparticles are stable. The sound speed of unparticles, $c_u^{2}$ in both regimes is given as \[
    c_{u}^{2}= 
\begin{dcases}
    \frac{y_0^3 \left(y_0^{\delta }-1\right)}{\delta \,
   (z+1)^3},& \text{if }  \mathcal{O}(10) \geq z \geq -1 \\
    1/3\,               & z \gg 1 \, ,
\end{dcases}
\]
where $y_0$ is the present-day value of $y$, which is very close to unity. Note that $ 0 < c_{u}^{2} < 1$ throughout the evolution. Hence, for the discussed range of parameters, $B<0,\,-3 <\delta<0$ perturbations are free from ghosts, gradient instability and superluminal propagation. 

\section{Consistency Condition}
\begin{figure}[] 
\centering
 \begin{adjustbox}{center}
\includegraphics[width=5.2cm, height=3.8cm]{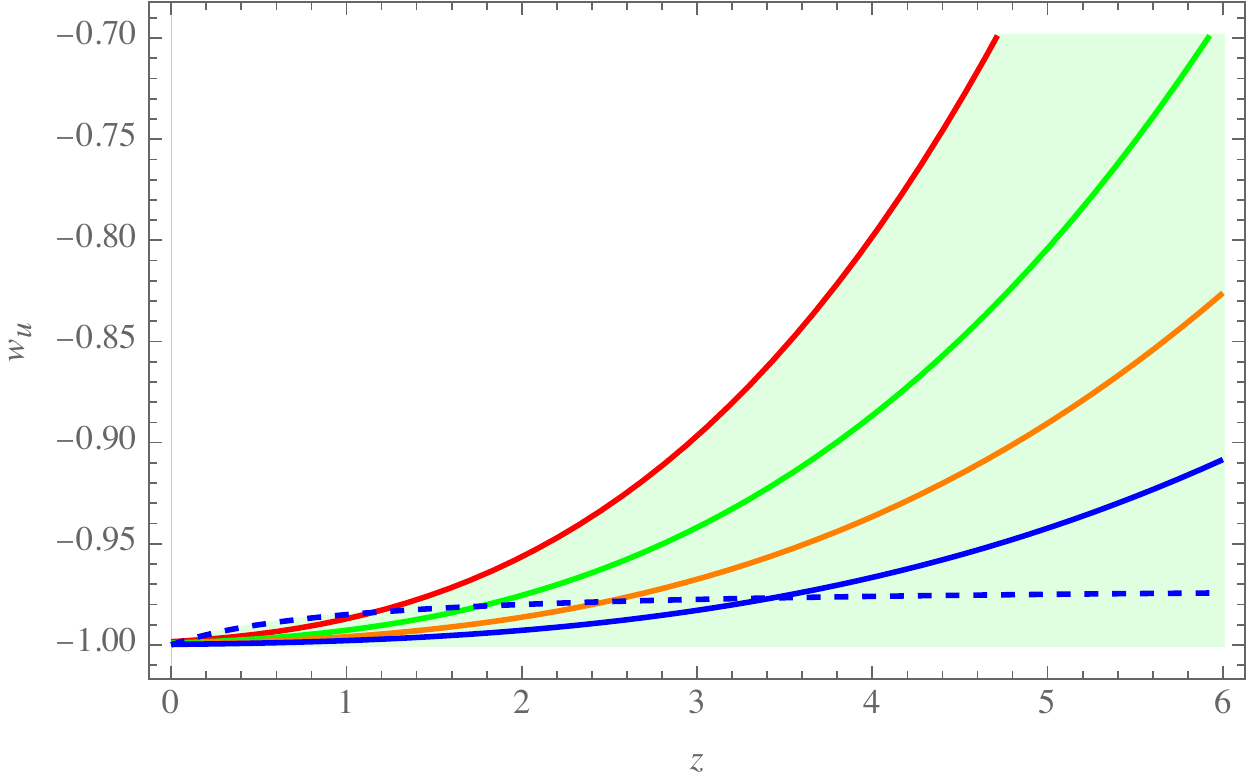}
\end{adjustbox}
\caption{\it The EOS of the unparticles DE model $w_u$ as a function of redshift. The lines correspond to $\Delta N_{eff}=0.19$ and $(-\delta=0.001,0.01,0.1,1)$ from top to bottom, respectively. For a given $\delta$ the shaded region below the graph represents the constraint of $\Delta N_{eff}\leq 0.19$. The dashed line corresponds to a CPL parameterization of $w=-1+0.03 z/(1+z)$.}
\label{fig:consistency}
\end{figure}

After establishing the stability of the unparticles DE model, let us discuss its predictions. Since the equation of state of unparticles $w_u$ naturally changes from that of radiation to that of a CC, it has observable effects both at early and late times. The equation of state of unparticles $w_u$, can be measured at late times $z\sim \mathcal{O}(few)$. The redshift dependence is rather different than the standard CPL parameterization \cite{Chevallier:2000qy,Linder:2002et}, making the model an easy target for detection. At early times, unparticles contribute to the number of relativistic degrees of freedom at CMB decoupling (and BBN), $\Delta N_{eff}$,
\be
\frac{\rho_u}{\rho_r}\simeq \frac{\Omega_{u0}}{\Omega_{r0}} 3(\delta+4)(-\delta)^{1/3}(y_0-1)^{4/3}=\frac{7}{8}\left(\frac{4}{11}\right)^{4/3}\Delta N_{eff} \label{eq:Neff}
\ee
where $\Omega_{r0}$ and $\Omega_{u0}$ are the present day relative densities of radiation and unparticles respectively, assuming $\rho_u$ accounts for all the DE. According to present data $\Delta N_{eff}\leq 0.19$ at $68\%$ confidence level \cite{Cooke:2013cba,Artymowski:2017pua,Ben-Dayan:2019gll}. One can use Eq. \eqref{eq:Neff} to find an upper bound on $y_0$, which gives $y_0-1 \lesssim 10^{-4.5}$. Similarly, at low redshift the equation of state of unparticles can be approximated as 
 \be 
  w_{u} \simeq  -1 + 4 (\delta +4) \left(y_0-1\right) (1+z)^3 \, , \label{eq:approxwu}
 \ee
 Substituting \eqref{eq:Neff} into \eqref{eq:approxwu} we can remove the dependence on the temperature of unparticles and get a consistency relation:
 \be \label{eq:consistency} 
 w_u(z)\simeq-1+0.58\,(1+z)^3\left(-1-\frac{4}{\delta}\right)^{1/4}\left[\frac{\Omega_{r0}}{\Omega_{u0}}\Delta N_{eff}\right]^{3/4}
 \ee
Figure \ref{fig:consistency} shows the equation of state as a function of redshift $w_u(z)$ up to redshift $z=6$ as expected from the Megamapper proposal \cite{Schlegel:2019eqc,Sailer:2021yzm}, after setting $\Omega_{u0}=0.6911,\, \Omega_{r0}=8.97 \times 10^{-5} $ \cite{Aghanim:2018eyx}. The different lines correspond to $\Delta N_{eff}=0.19$ and $(-\delta=0.001,0.01,0.1,1)$ from top to bottom, respectively. For a given $\delta$ the shaded region below the graph represents the constraint of $\Delta N_{eff}\leq 0.19$. The dashed line corresponds to a CPL parameterization of $w=-1+0.03 z/(1+z)$ which is allowed by Planck Observations. Notice the enhanced redshift dependence of the unparticle DE model.

\section{Unparticles DE model and Supernovae Data}
To demonstrate the possibilities of our model, we consider the publicly available Pantheon data \cite{Pan-STARRS1:2017jku} with $1048$ supernovae. Very low-redshift supernovae have significant peculiar velocities that may bias the inferred value of $H_0$, \cite{Ben-Dayan:2014swa} so we limit ourselves to supernovae at $z\geq 0.03$, leaving $953$ supernovae in the analysis. We do not account for other systematic errors \cite{Ben-Dayan:2012uam,Ben-Dayan:2012ccq,Ben-Dayan:2012lcv,Ben-Dayan:2013nkf,Ben-Dayan:2013eza,Ben-Dayan:2014iya,Ben-Dayan:2015zha}, so our analysis is mostly a proof of concept.
While there are good reasons to split the sample into low redshift and high redshift and fit $H_0$ and $\Omega_{m0}$ separately, we think a simultaneous fit is desirable as well, as it is equivalent to the parameter inference used in CMB analyses and better portrays the tension between the different measurements.\footnote{We thank Tzvi Piran for raising this point.} Considering the apparent magnitude given by
\begin{eqnarray}
\mu = 5 \,Log_{10}\, d_{L}(H_0\,,\Omega_{m_{0}},\delta,\Delta\, N_{eff}) + 25 + M_B,
\end{eqnarray}
where $d_{L}$ is the luminosity distance, and $M_B$ the absolute magnitude. 

\begin{figure}[h]
\centering
 \begin{adjustbox}{center}
\includegraphics[scale=0.5]{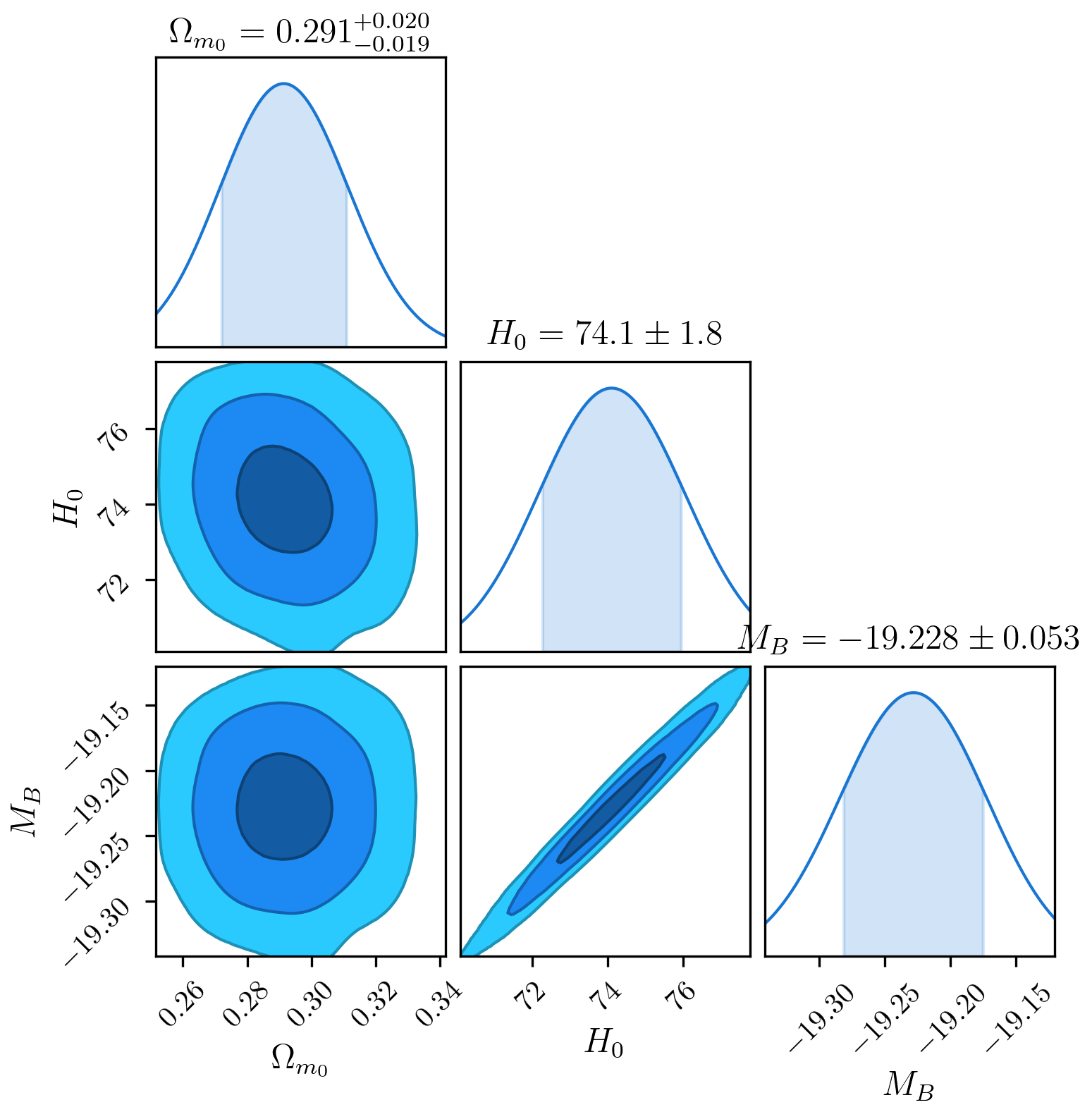} 
\end{adjustbox}
\caption{\it Likelihood contours and posterior distribution of parameters of $\Lambda$CDM using the Pantheon data set with a Gaussian prior on $M_B$.}
\label{fig:likeLCDM}
\end{figure}

We perform likelihood analysis to constrain the model parameters, using flat priors on the cosmological parameters and a Gaussian prior on $M_B=-19.23\pm 0.04$, \cite{Camarena:2019moy}. To implement the analysis, we use the public package MultiNest \cite{Feroz:2013hea} through the python interface PyMultiNest \cite{Buchner:2014nha}. 
First we analyzed the $\Lambda$CDM model. The results are presented in figure \ref{fig:likeLCDM}, and to a good accuracy reproduce the recent results reported in the literature \cite{Pan-STARRS1:2017jku,Buchner:2014nha}. We then performed a similar analysis for the unparticles DE model. 

\begin{figure}[h]  
 \begin{adjustbox}{center}
\includegraphics[scale=0.65]{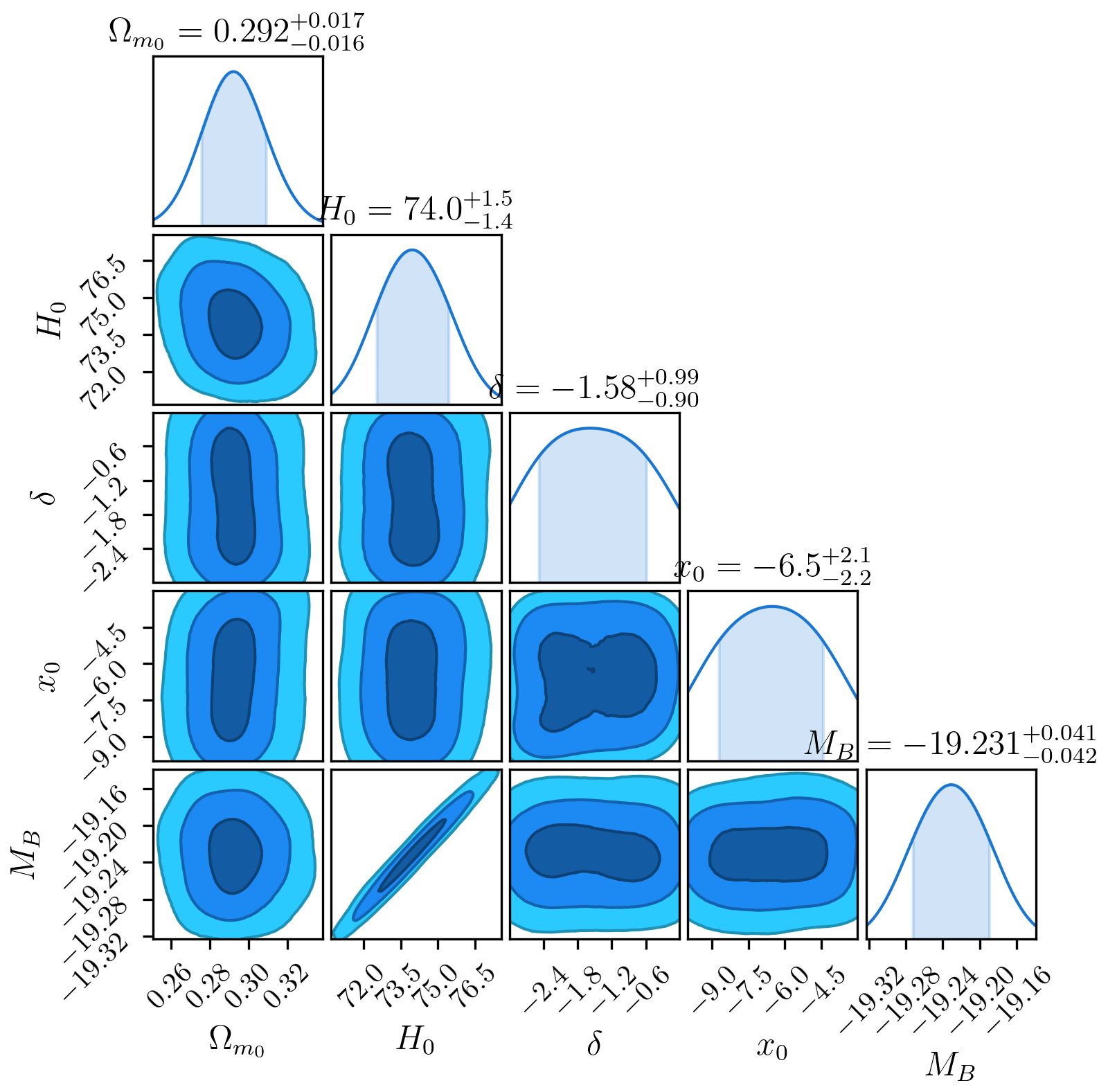} 
\end{adjustbox}
\caption{\it 
Same as figure \ref{fig:likeLCDM} with the parameters of unparticles DE model $\Omega_{m0},H_0,\delta$ and $x_0$.}
\label{fig:unpx0-dM}
\end{figure}

\begin{figure}[h]
\centering
 \begin{adjustbox}{center}
\includegraphics[scale=0.65]{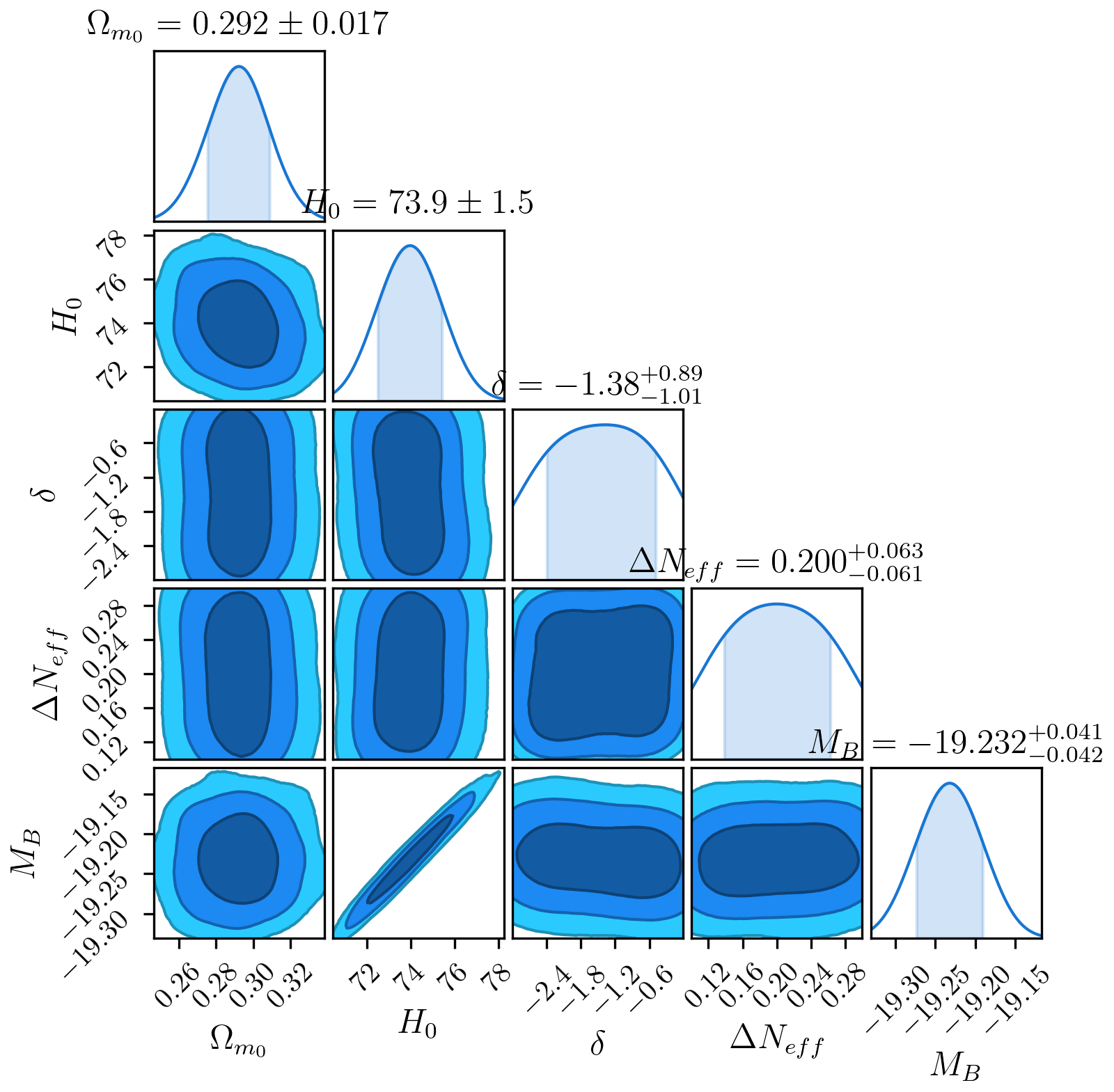} 
\end{adjustbox}
\caption{\it Same as figure \ref{fig:likeLCDM} with the parameters of unparticles DE model $\Omega_{m0},H_0,\delta$ and $\Delta N_{eff}$ using the consistency equation.}
\label{fig:unpNeff-dM}
\end{figure}

In figure \ref{fig:unpx0-dM} we show the results for $\Omega_{m0},H_0,\delta$ and $x_0=\log_{10}(y_0-1)$. As expected from the analytical formulae, Cosmology is rather insensitive to the value of $\delta$ except the case of $|\delta|\ll 1$. We do get some constraints on the temperature of unparticles today.
As explained this can be traded off for constraining $\Delta N_{eff}$ via the consistency condition \eqref{eq:consistency}, which is demonstrated in figure \ref{fig:unpNeff-dM}. 

\begin{figure}[h] \label{fig:unpNefffixdeltaM}
\centering
 \begin{adjustbox}{center}
\includegraphics[scale=0.55]{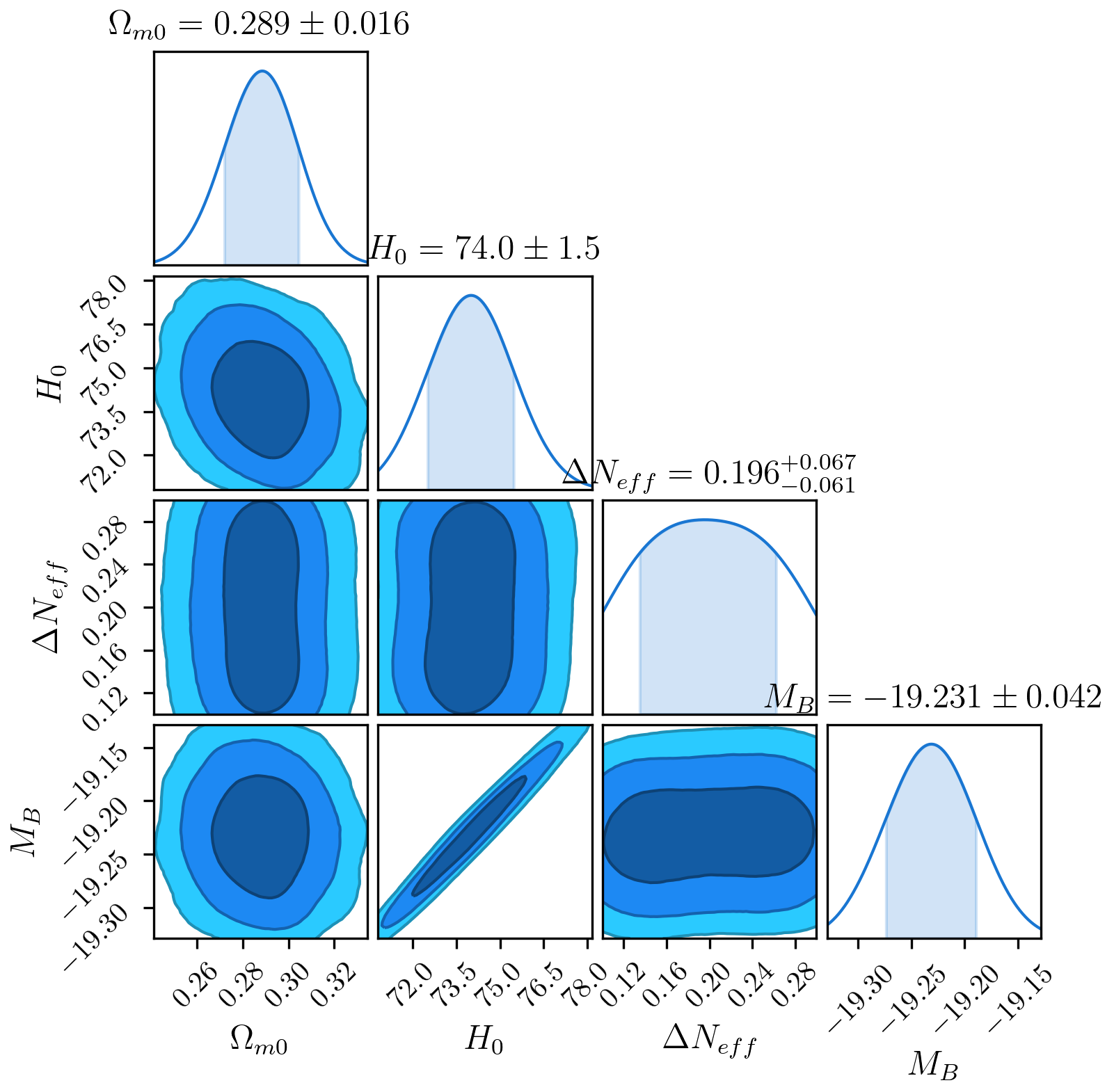} 
\end{adjustbox}
\caption{\it 
Same as figure \ref{fig:likeLCDM} with the parameters of unparticles DE model $\Omega_{m0},H_0,\delta = -0.001$ and $\Delta N_{eff}$ using the consistency equation.
}
\label{fig:unpNefffixdeltaM1}
\end{figure}

Given the weak dependence on $\delta$ we fix it to some value and repeat the analysis. Figure \ref{fig:unpNefffixdeltaM} shows the likelihood contours 
for $\delta=-0.001$, and figure \ref{fig:unpNefffixdeltaM1} corresponds to fixing $\delta=-1$. Notice that the unparticles DE model prefers a larger $\Delta N_{eff}$, and induces larger error bars in $H_0$. If $\Delta N_{eff} \simeq 0.2$, then CMB parameter estimation
taken at face value should yield $H_0\simeq 70 km/sec/Mpc$. 

\begin{figure}[h]
 \begin{adjustbox}{center}
\includegraphics[scale=0.55]{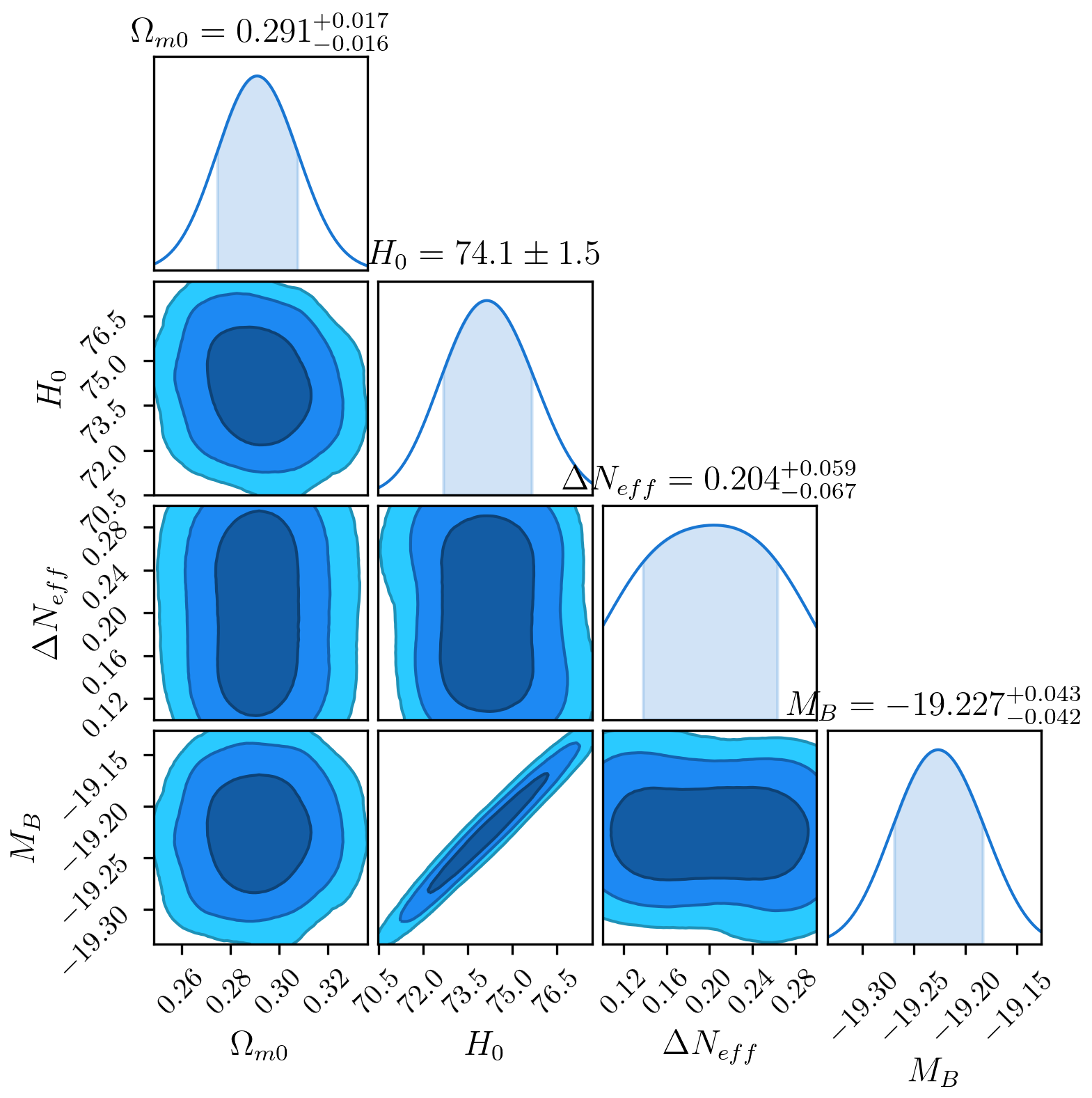} 
\end{adjustbox}
\caption{\it 
Same as figure \ref{fig:likeLCDM} with the parameters of unparticles DE model $\Omega_{m0},H_0,\delta = -1$ and $\Delta N_{eff}$ using the consistency equation.
}
\label{fig:likeUnpd2}
\end{figure}

\section{Conclusions}

In this work we compared our DE model presented in \cite{Artymowski:2020zwy} to the analysis of unparticle DE done in \cite{Abchouyeh:2021wey}. We showed that the allowed parameter space discussed by the Authors should be further restricted to $B>0$ and $\delta>-3$ if one wants to obtain unparticles with positive energy density, lack of NEC violation, and $da/dT<0$ in the $T\to 0$ regime.

Furthermore, following the results of \cite{Artymowski:2020zwy} we showed that for $B<0$ and $\delta \in (-3,0)$ the temperature of unparticles has a lower/upper bound dubbed $T_c$, which sets the scale of DE. In such a case, starting from $T>T_c$, one cannot reach $T\to0$ region via physical evolution of the Universe. 
Thus, for the significant part of the parameter space one obtains a valid DE solution, which prevents unparticles from ever reaching temperatures lower than $T_c$. This conclusion may only be modified if some higher order corrections to the beta function of unparticles modify equation \eqref{eq:basic}.

In addition, we have proven that the realistic solutions for DE remain stable, with an attractor solution $\Omega_u\to1$ regardless of initial conditions as long as the initial temperature of unparticles is larger than $T_c$. We derived a consistency relation between the maximal allowed contribution of unparticle energy density to the total energy of the Universe in the early Universe $\Delta N_{eff}$ with the present-day value of unparticle equation of state $w_u$, that may be detectable with near future observations. Finally, we fitted the model to supernovae data. 
For the unparticles emergent DE model we derived the first constraints using SN data weakly favoring positive $\Delta N_{eff}$. The expected Hubble parameter is similar to the $\Lambda$CDM value of our analysis and other reported analyses in the literature. Hence, any improvement on the Hubble tension should come from reanalyzing the CMB data and validating the analytical estimate. 

In brief, the unparticles emergent DE model is promising and we intend to perform a likelihood analysis taking into account additional cosmological observations most notably CMB to further assess the validity of the model. 

\textbf{Acknowledgments:}{ We thank Jakob Robnik for useful discussions. The work was partly funded by project RA1800000062.} This work has been supported by the National Science Centre (NCN),
Poland, through grant No. 2021/41/B/ST10/00823.

\end{document}